\documentclass{article}
\usepackage{graphicx, braket, amsmath, cite, amssymb, subcaption, tikz}
\usetikzlibrary{positioning,arrows.meta}
\usepackage[colorlinks = true]{hyperref}
\usepackage[margin = 1in]{geometry}

\tikzset{
	mynode/.style={
		circle, draw=black!70, minimum size=9mm, font=\small, inner sep=0pt
	},
	edge/.style={black, line width=0.8pt},
	elabelstyle/.style={font=\scriptsize\color{black}, inner sep=0pt}
}

\newtheorem{definition}{Definition}

\newtheorem{procedure}{Procedure}

\DeclareMathOperator{\Nbd}{Nbd}

\title{Community detection in network using Szegedy quantum walk}
\author{Md Samsur Rahaman\thanks{Email: \texttt{msr165031@gmail.com}}, Supriyo Dutta\thanks{Email: \texttt{dosupriyo@gmail.com}} \\ \small{Department of Mathematics,} \\ \small{National Institute of Technology Agartala,} \\ \small{Jirania, West Tripura, India - 799046.}}
\date{} 

\begin{document}
	
	\maketitle
	
	\begin{abstract}
		In a network, the vertices with similar characteristics construct communities. The vertices in a community are well-connected. Detecting the communities in a network is a challenging and important problem in the theory of complex networks. One approach to solving this problem uses the classical random walks on graphs. In quantum computing, quantum walks are the quantum mechanical counterparts of classical random walks. In this article, we employ a variant of Szegedy's quantum walk to develop a procedure for discovering the communities in networks. The limiting probability distribution of quantum walks assists us in determining the inclusion of a vertex in a community. We apply our community detection procedure to a variety of graphs and social networks, including the relaxed caveman graph, $l$-partition graph, Karate club graph, and the dolphin's social network, among others. \\
		\noindent \textbf{Keywords}: Clustering, Community detection, Szegedy quantum walk, Complex networks.
	\end{abstract}
	
	\tableofcontents
	
	\section{Introduction}
	
	Networks are everywhere! Human beings build up networks among themselves for their progress and prosperity. They establish roads, rails, airlines, and the internet. They all are different networks \cite{estrada2012structure, newman2010networks}. Networks are inherent in nature. The human brain and protein-protein interaction network are two important examples of networks in nature \cite{albert2002statistical}. In a network, elements of similar characteristics are well-connected. It distributes all the elements in a network into communities. Detecting the communities in a given network is an important problem in network science \cite{girvan2002community, fortunato2010community, malliaros2013clustering, porter2009communities}. 
	
	Combinatorial graphs \cite{west2001introduction, gross2003handbook} form the mathematical foundation of networks. We use the words graphs and networks synonymously throughout the article. A graph is a combination of objects and connection between them, where the objects are represented by the vertices and connection are represented by edges between them. The real-world networks display big inhomogeneities in the distribution of edges. The concentrations of edges within special groups of vertices is high. We call them communities, or clusters. Due to the existence of connections between the vertices, we prefer to use the word community instead of cluster.
	
	Detecting the communities in a network is not an easy task. There are different methods of community detection discussed in the literature \cite{kernighan1970efficient, suaris2002algorithm, barnes1982algorithm, pothen1997graph, sapatinas2004elements, moghaddam2026differentially}. Many of the community detection algorithms are  computationally NP-hard, or NP-complete \cite{garey2002computers, matsuda1999classifying, drineas2004clustering}.  The random walks on graphs provide a crucial method to detect the communities \cite{zhou2003distance, zhou2004network, weinan2008optimal, dongen2000performance}. A random walker travels between the vertices following the connectivity structure of the graph. As the movement of the walker is random, at any time-instance, there is a probability distribution for getting the walker at different vertices. Therefore, we get a sequence of probability distributions that converges to a limiting probability distribution under certain conditions. Usually, the inter-community edges are higher than the probability of the intra-community edges. In this way, the limiting probability distribution of random walks assists us in detecting the communities.
	
	In quantum computing, the quantum walks  \cite{portugal2013quantum} are the quantize versions of classical random walks on the graphs \cite{watrous2001quantum, woess2000random}. The quantum walks are applied in quantum communication \cite{coutinho2014quantum, dutta2023quantum}, and computation \cite{lovett2010universal, paparo2013quantum, dutta2025discrete}. There are multiple proposals of quantum walks, such as discrete-time quantum walks \cite{higuchi2017periodicity}, continuous-time quantum walks \cite{solenov2006continuous}, open quantum walks \cite{attal2012open, rani2024non}, etc. The Szegedy quantum walks \cite{higuchi2017periodicity} converts the transition matrix of a classical random walk to a unitary matrix to govern the quantum walker on graphs. In addition we can define it on all the simple graph. Therefore, the Szegedy quantum walk is an efficient way to replace the idea of random walk for developing a quantum algorithm. 
	
	The graph clustering problem has recently been considered in the literature of quantum computing \cite{ushijima2017graph, pramanik2020quantum, volya2021quantum}. The quantum walks are also introduced for finding the clusters in graphs \cite{roy2017novel, mukai2020discrete}. Article \cite{mukai2020discrete} utilizes the discrete-time coined quantum walks to identify the communities in different real-world networks. The Fourier and Grover coins are used as the quantum coin operators. It is demonstrated that the Fourier coin quantum walks localize within communities, which effectively reveal the modular structures of networks. Article \cite{roy2017novel} improves the clustering speed and accuracy. Article \cite{volya2021quantum} discusses another efficient approach to clustering on directed and mixed graphs without using quantum walks.
	
	The motivation of this work comes from the Szegedy quantum walk on the Barbell graphs. A barbell graph has two communities of vertices that are joined to a central vertex by two inter-community edges. Ref. \autoref{barbell_graph}. We observed that the limiting probability distribution of Szegedy quantum walk from a particular initial state has higher probability values for the inter-community edges. Their removal generates the communities in the graph. This observation assists us in developing a method for community detection in networks.  
	
	In this article, we present a new method for finding the communities in networks, which is applicable to all simple graphs with communities. For our constructions, we apply the model of Szegedy quantum walks discussed in \cite{higuchi2017periodicity}. To the best of our knowledge, this is the first research article where Szegedy quantum walk is applied for community detection. The quantum walk on graphs provides a limiting probability distribution on the edges of the graph. It assists us in determining a vertex to include in a community. For the numerical experiments, we apply our community detection techniques on different networks which are already used in literature \cite{mukai2020discrete, fortunato2010community, zhang2013finding}. Our results also matches with them.	Due to the difference between the random walks and quantum walks, we observe that the communities are different in some cases.
	
	The structure of this article is as follows: The primary concepts are discussed in \autoref{priliminary_section}. In \autoref{Community_detection_on_graphs}, we describe our procedures of community detection. The \autoref{Numerical_experiments} contains the communities in different networks generated by our procedure. Then we conclude the article. In Appendix we discuss the choice of our initial quantum state, and the convergence of the probability distribution.

	\section{Preliminary}\label{priliminary_section}
	
	\subsection{A brief Overview of graph theory}\label{A_brief_Overview_of_graph_theory}
	
	\label{A_brief_Overview_of_graph_theory}
	A graph $G=(V(G),E(G))$ consists of a set of vertices $V(G)$ and a set of edges $E(G) \subseteq V(G) \times V(G)$ \cite{west2001introduction}. Throughout this article, $n$ and $m$ denote the number of vertices and edges, respectively. An undirected edge between the vertices $u$ and $v$ is denoted by $(u, v)$. A loop is an edge joining a vertex with itself. A graph with the undirected edges, and without loops is called a simple graph. We do not consider weighted graphs, in this article. A graph with $n$ vertices is said to be a complete graph, $K_n$ if there is an edge between any two distinct vertices.
	
	In a simple graph $G$, two vertices $u$ and $v$ are said to be adjacent if there is an edge $(u, v)$. Also, the edge $(u, v)$ is incident to the vertices $u$ and $v$. The number of incident edges to a vertex $v$ is called the degree of $v$, which is denoted by $d_G(v)$. The neighborhood of a vertex $v$ in $G$ is the set $N(v) = \{v: (u, v) \in E(G)\}$. A path between the vertices $v_0$ and $v_n$ is a sequence of vertices and edges $v_0, e_1, v_1, e_2, v_2, \dots e_n, v_n$ where $e_i = (v_{i - 1}, v_i)$ for $i = 1, 2, \dots n$, where no vertex is repeated. The number of edges in a path is called the length of a path. A graph is called connected if there is a path between any two vertices. The path between $u$ and $v$ with minimum length is called the shortest path.  A graph $G_1 = (V(G_1),E(G_1))$ is said to be a subgraph of a graph $G=(V(G),E(G))$, if $V(G_1) \subseteq V(G)$, and $E(G_1) \subseteq E(G)$. A subgraph $G_1$ of $G$ is called an induced subgraph if $E(G_1) = \{(u,v): (u,v) \in E(G) ~\text{and}~ u, v \in V(G_1)\}$.
	
	A directed edge from $u$ to $v$ is denoted by $\overrightarrow{(u, v)}$. All the edges in a directed graph $\overrightarrow{G} = (V(\overrightarrow{G}), E(\overrightarrow{G}))$ is directed. Given a directed edge $\overrightarrow{e} = \overrightarrow{(u, v)}$, the vertices  $u = o(\overrightarrow{e})$, and $v = t(\overrightarrow{e})$ are said to be the origin and the terminal of $\overrightarrow{e}$, respectively. The indegree of a vertex $v$ is denoted by $d^{(i)}_{\overrightarrow{G}}(v)$, which is the number of edges $\overrightarrow{e}$ such that $v = t(\overrightarrow{e})$. Also, the outdegree of a vertex $v$ is denoted by $d^{(o)}_{\overrightarrow{G}}(v)$, which is the number of edges $\overrightarrow{e}$ such that $v = o(\overrightarrow{e})$. The inverse of a directed edge $\overrightarrow{e} = \overrightarrow{(u, v)}$, is $\overrightarrow{e}^{-1} = \overrightarrow{(v, u)}$. Therefore, $o(\overrightarrow{e}) = t(\overrightarrow{e}^{-1})$ and $t(\overrightarrow{e}) = o(\overrightarrow{e}^{-1})$.
	
	In this article, we assume that a vertex belongs to at most one community. The set of communities $\mathcal{C} = \{C_1, C_2, \dots, C_k \}$ in a graph is a partition on $V(G)$, such that, $\cup_{i = 1}^n C_i = V(G)$ and $C_i \cap C_j = \emptyset$ for $i \neq j$. We say that $\mathcal{C}$ is trivial if either $k = 1$, or $C_i$ is a singleton set for all $i$. We identify a community $C_i$ with the induced subgraph $G[C_i] = (C_i, E(C_i))$ of the graph $G$ for every community $C_i$, where $E(C_i) := \{(v, w) \in E : v, w \in C_i \}$. The $E(\mathcal{C}) = \cup_{i = 1}^k E(C_i)$ is the set of intra-community edges induced by $\mathcal{C}$. The number of edges in $E(\mathcal{C})$ is denoted by $m(\mathcal{C})$. Also, $E(G) - E(C) = \{(u, v) : u \in C_i,\ v \in C_j,\ i \neq j\}$ is the set of inter-community edges of $G$ induced by $\mathcal{C}$. The number of intra-community edges is more than the number of inter-community edges.

	The community detection in graph is the task of grouping the vertices of the graph into the communities. Recall that, different procedures of community detection develops different communities of vertices in a graph. There are a number of measures to quantify the efficiency of a community detecting procedure. Recall that $\mathcal{C}$ and $m$ are the intra-community edges, and the edges in a graph $G$, respectively. The coverage is one measure to quantity, which is defined by $\operatorname{Coverage}(\mathcal{C}) = \frac{m(\mathcal{C})}{m}$. The large value of $\operatorname{Coverage}(\mathcal{C})$ indicates the better quality of $\mathcal{C}$. In classical computing the graph community detection problem is an NP-hard problem when $k \geq 3$ \cite{brandes2003experiments}.
	
	Communities in a network depends on the connectivity properties of a network. Not all graphs preserve a structure of community in it. For examples, the complete graph with any number of vertices results a single community. Well-known graphs with community structure includes different varieties of caveman graphs, planted partition graphs, as well as different complex networks such as the Zachary's karate club graph, etc.

	\subsection{An introduction to quantum walks}
	
	From a connected simple graph $G = (V(G), E(G))$ we construct a directed graph $\overrightarrow{G}$ by replacing every undirected edge $(u, v)$ by two directed edges $\overrightarrow{e} = \overrightarrow{(u, v)}$ and $\overrightarrow{e^{-1}} = \overrightarrow{(v, u)}$. Let the edges $\overrightarrow{e_{1}}, \overrightarrow{e_{2}}, \dots, \overrightarrow{e_{m}}, \overrightarrow{e_1^{-1}}, \overrightarrow{e_2^{-1}}, \dots, \overrightarrow{e_m^{-1}}$ in $\overrightarrow{G}$ are marked by the integers $0, 1, 2, \dots (2m -1)$. The construction of $\overrightarrow{G}$ from $G$ indicates that $d_G(v) = d^{(o)}_{\overrightarrow{G}}(v) = d^{(i)}_{\overrightarrow{G}}(v)$.
	
	At time $t$, let the walker is at a vertex $v$. At time $(t + 1)$, following one of the $d_v^{(o)}$ outgoing edges it reaches to a neighboring vertex. The walker chooses one of the out-going edges with probability $p(\overrightarrow{e}) = \frac{1}{d^{(o)}_v} = \frac{1}{d_v}$, where $v = o(\overrightarrow{e})$. Therefore, $\sum_{\overrightarrow{e} : o(\overrightarrow{e}) = v} p(\overrightarrow{e}) = 1$, for each vertex $v\in V$.  
	
	Corresponding to the directed graph $\overrightarrow{G}$ with $2m$ edges, we consider a Hilbert space $\mathcal{H}^{2m}$ spanned by $\left\{ |0\rangle, |1\rangle, |2\rangle, \dots, |2m - 1\rangle \right\}$, where $\ket{i} = (0, 0, \dots 1 ((i+1)\text{-th position)}, 0, \dots, 0)^\dagger$ corresponds the $i$-th directed edges, for $i = 0, 1, \dots (2m -1)$. Clearly, for any directed edge $\overrightarrow{(i,j)}$ in $E(\overrightarrow{G})$, there exists an unit vector $\ket{\overrightarrow{(i,j)}}$ in the computational basis of $\mathcal{H}^{2m}$. The Szegedy's quantum walk is determined by a unitary operator $U_{sz} = (u_{\overrightarrow{e}, \overrightarrow{f}})_{2m \times 2m}$, where 
	\begin{equation}
		\label{Szegedy matrix}
		u_{\overrightarrow{e}, \overrightarrow{f}} = 
		\begin{cases} 
			2 \sqrt{p(\overrightarrow{e}) p(\overrightarrow{f^{-1}})} - \delta_{(\overrightarrow{e}, \overrightarrow{f^{-1}})} & \text{if}~ t(\overrightarrow{f}) = o(\overrightarrow{e}); \\
			0, & \text{otherwise}.
		\end{cases}
	\end{equation}
	
	Here, $\delta_{(\overrightarrow{e}, \overrightarrow{f^{-1}})}$ is the Kronecker delta, which is defined by \begin{equation}
		\delta_{(\overrightarrow{e}, \overrightarrow{f^{-1}})} = 
		\begin{cases} 
			1 & \text{if}~ \overrightarrow{e} = \overrightarrow{f^{-1}}; \\
			0 & \text{otherwise}.
		\end{cases}
	\end{equation}
	Let $\ket{\psi_t}$ represents the state of the walker at time $t$. For $t = 0, 1, 2, \dots$, the state of the walker at time $(t + 1)$ is represented by $\ket{\psi_{(t + 1)}} = U_{sz} \ket{\psi_t}$.
	
	In this article, the walker begins its journey with a particular initial state. Let $V_{\max}$ is the set of vertices with maximum degree in the graph $G$ and $E_{\operatorname{init}} = \{\overrightarrow{(i,j)} \mid i \in V_{\max}\} \subset E(\overrightarrow{G})$ is the set of all outgoing edges from the vertices in $V_{\max}$. The initial state is determined by
	\begin{equation}\label{initial_state}
		\ket{\psi_0} = \frac{1}{\sqrt{|E_{\operatorname{init}}|}} \sum_{i \in V_{\max}}\ket{\overrightarrow{(i,j)}}.
	\end{equation}
	The procedure of community detection depends on the selection of the initial state, which is further explored in \autoref{Dependency_on_initial_state}. At  $t\textsuperscript{th}$ step of quantum walk for $t = 0, 1, 2, \dots$, the state vector of the walker is
	\begin{equation}
		\ket{\psi_t} = U_{sz}^t \ket{\psi_0} = |\psi_t\rangle = \sum_{k=0}^{2m-1} a_{t, k} |k\rangle,
	\end{equation} 
	where $\ket{k}$ is the basis vector of $\mathcal{H}^{2m}$ corresponding to the directed edge $\overrightarrow{k}$. Also, $a_{t, k}$ are amplitudes, satisfying the condition 
	\begin{equation}\label{eq:probability}
		\sum_{k=0}^{2m-1} |a_{t, k}|^2 = 1, ~\text{for all}~ t.
	\end{equation}

	\section{Community detection on graphs}
	\label{Community_detection_on_graphs} 
	
	Let $P_k = \ket{k}\bra{k}$ be the projection operator corresponding to the basis vector $\ket{k}$. Then, the probability of getting the walker on the $k$-th edge for $0 \leq k \leq 2m - 1$ at time $t$ is 	
	\begin{equation}\label{edge_probability_at_t_th_step}
		p_t(\overrightarrow{k}) = \braket{\psi^\dagger | P_k| \psi} = |\langle k | \psi \rangle|^2 = |\alpha_{t, k}|^2.
	\end{equation}
	Using equation (\ref{edge_probability_at_t_th_step}), we construct an edge probability vector for $\overrightarrow{G}$ at the $t$-th step of quantum walk as
	\begin{equation}
		\begin{aligned}
			\overrightarrow{P}_t &=
			\bigl(
			p_t(\overrightarrow{e}_1), \ldots, p_t(\overrightarrow{e}_m), p_t(\overrightarrow{e}_1^{-1}), \ldots, p_t(\overrightarrow{e}_m^{-1})
			\bigr).
		\end{aligned}
	\end{equation}
	Note that, every edge $e$ in $G$ contributes two directed edges $\overrightarrow{e}$ and $\overrightarrow{e^{-1}}$, in $\overrightarrow{G}$. Thus, we determine the probability of $e$ in $G$ at time $t$ as $p_t(e) = p_t(\overrightarrow{e}) +  p_t(\overrightarrow{e^{-1}})$. Therefore, the edge probability vector of $G$ at time $t$ as
	\begin{equation}
		P_t = (p_t(e_1), p_t(e_2), \dots, p_t(e_m)),
	\end{equation}
	where $p_t(e_i) = p_t(\overrightarrow{e_i}) + p_t(\overrightarrow{e_i^{-1}})$. Note that, due to the unitary dynamics, $P_t$ need not converge to a limiting distribution as $t \rightarrow \infty$ \cite{ambainis2003quantum}. But, the time average probability distribution converges. Hence, we have the following definitions.
	\begin{definition}
		\label{edge_probability_vectors}
		The edge-probability vector at time $T$ of a graph $G$ is defined by
		\begin{equation*}
			\begin{split}
				& \Pi^{(T)} 
				= \bigl(\pi_1^{(T)}, \pi_2^{(T)}, \ldots, \pi_m^{(T)}\bigr) = \frac{1}{T+1}\sum_{t=0}^{T} P_t \\
				&= \frac{1}{T+1}
				\bigl(\sum_{t=0}^{T} p_t(e_1), \sum_{t=0}^{T} p_t(e_2), \ldots, \sum_{t=0}^{T} p_t(e_m)\bigr),
			\end{split}
		\end{equation*}
		where $\pi_k^{(T)} = \frac{1}{T+1}\sum_{t=0}^{T} p_t(e_k)$ for $k = 1, 2, \ldots, m$.
	\end{definition}
	
	Note that, the edge probability vector proceeds to a limiting distribution when $T \rightarrow \infty$, which we define as the edge probability vector of a graph.
	\begin{definition}
		\label{limiting_edge_probability_distribution}
		For a graph $G$ the edge-probability vector 
		\begin{equation*}
			\begin{split}
				& \Pi
				= \bigl(p(e_1), p(e_2), \ldots, p(e_m)\bigr) = \lim_{T \to \infty} \frac{1}{T+1}\sum_{t=0}^{T} P_t \\
				&= \lim_{T \to \infty} \frac{1}{T+1}
				\Bigl(
				\sum_{t=0}^{T} p_t(e_1), \sum_{t=0}^{T} p_t(e_2), \ldots, \sum_{t=0}^{T} p_t(e_m)
				\Bigr).
			\end{split}
		\end{equation*}
	\end{definition}
	
\begin{figure*}[t]
		\centering
		\setlength{\abovecaptionskip}{1pt}
		\setlength{\belowcaptionskip}{0pt}
		
		\begin{subfigure}[b]{0.32\textwidth}
			\includegraphics[width=\linewidth,height=2.6cm]{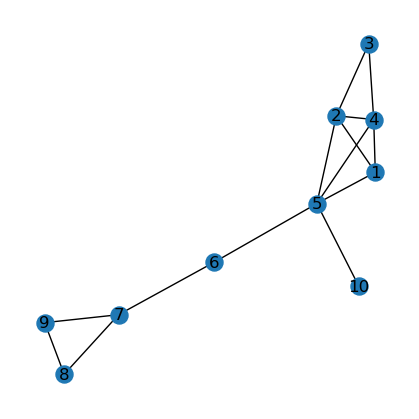}
			\caption{Graph $G$}
			\label{Random_graph}
		\end{subfigure}
		\hfill
		\begin{subfigure}[b]{0.32\textwidth}
			\includegraphics[width=\linewidth,height=2.6cm]{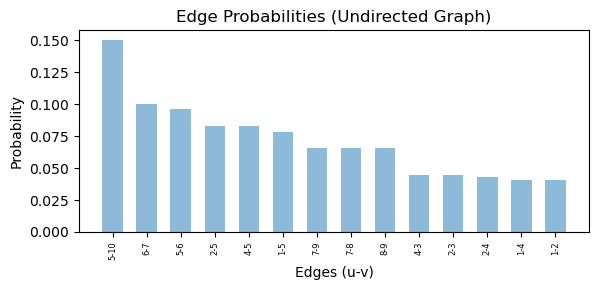}
			\caption{Edge probability in the limiting distribution $\Pi$ is depicted as bar diagram.}
			\label{edge_probability_diagram}
		\end{subfigure}
		\hfill
		\begin{subfigure}[b]{0.32\textwidth}
			\includegraphics[width=\linewidth,height=2.6cm]{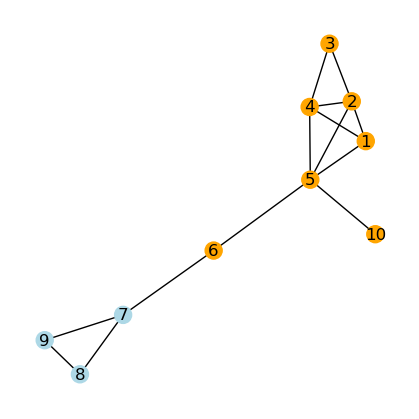}
			\caption{The vertex $5$ in $G$ has the maximum degree. The green vertices are in $\Nbd(5)$, which is generated following Procedure \ref{Main_procedure}.}
		\end{subfigure}
		
		\vspace{1mm}
		
		\begin{subfigure}[b]{0.32\textwidth}
			\includegraphics[width=\linewidth,height=2.6cm]{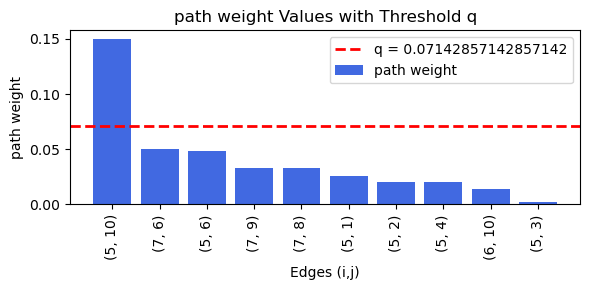}
			\caption{The path weights of the paths  between different vertices in $G$ are calculated by equation (\ref{path_weight}). They are plotted using bar diagram.}
		\end{subfigure}
		\hfill
		\begin{subfigure}[b]{0.32\textwidth}
			\includegraphics[width=\linewidth,height=2.6cm]{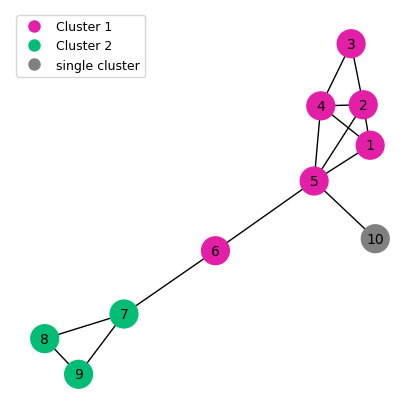}
			\caption{Communities in $G$ are extracted without refinement. Communities are represented by three different colors.}
		\end{subfigure}
		\hfill
		\begin{subfigure}[b]{0.32\textwidth}
			\includegraphics[width=\linewidth,height=2.6cm]{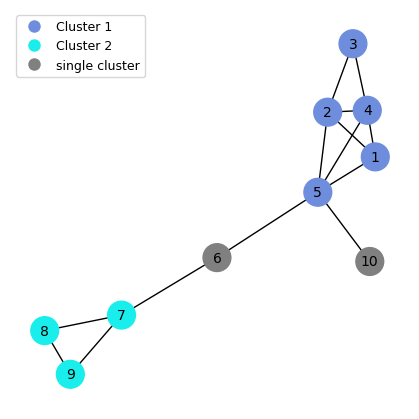}
			\caption{Communities in $G$ are extracted with refinement.Communities are represented by three different colors.}
		\end{subfigure}
		\caption{Different steps for generating communities on the graph $G$, depicted in Sub-figure  \ref{Random_graph}, are presented.}
		\label{Example_of_clustering}
	\end{figure*}

	For our computation, rather than calculating $ T \to \infty$, we find $\frac{1}{T + 1}\sum_{t = 0}^{T} P_t$ for a  finite $T$. We apply the standard Euclidean distance to differentiate two edge probability vectors $\Pi^{(T)}$ and $\Pi^{(T - 1)}$, which is
	\begin{equation}
		\begin{aligned}
			\left\lVert \Pi^{(T)} - \Pi^{(T-1)} \right\rVert
			&= \left\lVert \frac{1}{T+1}\sum_{t=0}^{T} P_t
			- \frac{1}{T}\sum_{t=0}^{T-1} P_t \right\rVert. \\[3pt]
		\end{aligned}
	\end{equation}
	To work out the edge probability vector of $G$, we compute a sequence of edge probability vectors $\{\Pi^{(t)}\}_{t = 0}^T$, where 
	\begin{equation}\label{stopping _criterion}
		\left|\Pi^{(T)} - \Pi^{(T - 1)}\right| < \epsilon.
	\end{equation}
	In our numerical experiments, $\epsilon = 0.0001$, which  assists us to find the accurate value of $\Pi$ to three decimal places. In \autoref{edge_probability_diagram}, we present the edge-probabilities of the graph in \autoref{Random_graph}. The convergence of the probability vectors are discussed in Appendix.
	
	Now, we include a few graph-theoretic ideas. A vertex with degree $1$ is adjacent to only one vertex, which may belong to a community. A vertex with degree two lies on a path between two neighboring vertices. These three vertices may form a single community, or the neighboring vertices may belong to two different communities. But, a vertex with greater degree is adjacent to many other vertices. Thus, it belongs to only one community. It insists us to arrange the vertices in a descending order of their degrees. In every step, we consider the highest degree vertex among the vertices which are not allotted to a community.
	
	There may be a number of vertices in a community, which are not an immediate neighbor of the vertex with highest degree in it. Also, there are non-neighboring vertices which are not in the community. When a vertex is in the community we observe that most of its neighbours are also in the same community. Therefore, we need a condition for including a given vertex in a particular community. Recall that $N(v_i) = \{v: (u, v) \in E(G)\}$. Let $\Nbd(v_i)$ be the neighborhood of $v_i$ in the community of vertices containing $v_i$. If $|N(v_j) \cap \Nbd(v_i)| \geq \frac{|N(v_j)|}{2}$ holds, we assume that $v_j$ is in the community of $v_i$.
	
	The final selection of a vertex in a community depends on the edge probability vector $\Pi$. We determine the shortest path between $v_i$ and $v_j$. The walker follows the path means it follows all the consecutive edges. As there is no non-Markovian componenet in our quantum walk, the probability of getting the walker on the edges of a path is independent to each other. Therefore, the probability of getting the walker on a path is the product of the probabilities. It leads us to define the path weight between two vertices. 
	
	\begin{definition}\label{path_weight}
		Let the shortest path between $u$ and $v$ in graph $G$ be $u = u_0, e_1, u_1, e_2, u_2, \dots e_n, u_n = v$. We define the path weight between the vertices $u$ and $v$ by
		\begin{equation}
			PW(u, v) = \frac{1}{d_G(v)} \prod_{i=1}^{n} p(e_i),
		\end{equation}
		where $p(e_i)$ is the probability corresponding to the edge $e_i$ in the edge probability vector $\Pi$.
	\end{definition}

	Let $v_i$ be a vertex of maximum degree among the vertices which are not included in a community. The community consists of $v_i$ is denoted by $C(v_i)$. Initially, $C(v_i) = \{v_i\}$. The Procedure mentioned below add the neighboring vertices of $v_i$ in $C(v_i)$.
	\begin{procedure}\label{Main_procedure}
		Follow the steps mentioned below:
		\begin{enumerate}
			\item 
			Arrange the vertices of $G$ as $V = \{v_1, v_2, \dots v_n\}$, where $d_G(v_i) \geq d_G(v_{i + 1})$. 
			
			\item 
			Repeat the following steps if $V \neq \emptyset$:
			
			\begin{enumerate}
				\item 
				Consider the first element of $V$, say $v_i$.
				\item 
				Construct $\Nbd(v_i) = N(v_i)$.
				\item 
				Let $v_j \notin N(v_i)$ is a vertex satisfying 
				\begin{equation}\label{nbd_criterion}
					|N(v_j) \cap \Nbd(v_i)| \geq \frac{|N(v_j)|}{2},
				\end{equation}
				then add $v_j$ in $\Nbd(v_i)$.
				\item 
				Construct the community with $v_i$, by setting $C(v_i) = \{v_i\}$.
				\item 
				Add all $v_j \in \Nbd(v_i)$ in $C(v_i)$ if $PW(v_i, v_j) < q$.
				\item 
				Update the set $V$ by $V - C(v_i)$.
			\end{enumerate}
			
			\item 
			Return the communities of $C(v_i)$.
		\end{enumerate}
	\end{procedure}	
	
	In \autoref{Main_procedure} the quantum walk is useful to calculate the probability values in finding $PW(v_i, v_j)$ mentioned in \autoref{path_weight}. Also, $q$ is a parameter. If $q = 0$ then the number of communities is equal to the number of vertices in the graph. When $q > \max_{(u, v) \in E(G)}\{PW(u, v)\}$ then $C(v_i) = \Nbd(v_i)$.
	
	In \autoref{path_weight}, we multiply $\frac{1}{d_G(v)}$ to normalize node influence. It provides more preference to a high-degree node to be added in $C(v_i)$, because a node with higher degree reduces the product $\prod_{i=1}^{n} p(e_i)$ than a vertex with lower degree. 
	
	After constructing a community $C(v_i)$ using Procedure \autoref{Main_procedure}, it may be observed that a number of additional vertices are in the community. For example, consider \autoref{Example_of_clustering}. Here, the vertex $6$ is added in  the community $C(1)$. But the structure of the graph suggests that vertex $6$ should form a separate community. Therefore, need to remove it from $C(1)$. To discard those vertices we perform a refinement using the below procedure. 
	
	\begin{procedure}
		\label{refinement_procedure}
		Let a community $C(v_i)$ is created using Step $1$ and $2$ of \autoref{Main_procedure}. Apply the following step before proceeding to step $3$.
		\begin{enumerate}
			\item
			For any vertex $v_j \in C(v_i)$ if $d_{C(v_i)}(v_j) \leq \frac{d_G(v_j)}{2}$ then remove $v_j$ from $C(v_i)$. 
		\end{enumerate}
	\end{procedure}
	
	Therefore, our method of community detection is a combination of the following steps, mentioned in the below procedure.
	\begin{procedure}\label{combined_procedure}
		To find the communities in a simple graph $G$ perform the following steps:
		\begin{enumerate}
			\item 
			\textbf{Generation of a directed graph:} Generate a directed graph $\overrightarrow{G}$ from $G$ by assigning opposite orientations on every undirected edge. 
			\item 
			\textbf{Construction of initial state:} Corresponding to every edge $i$ assign a state $\ket{i}$ of computational basis of dimension $2m$. Construct the initial state $\ket{\psi_0}$ using equation (\ref{initial_state}).
			\item 
			\textbf{Application of quantum walk:} Construct the Szegedy matrix using equation (\ref{Szegedy matrix}) and apply it to the initial state $\ket{\psi_0}$, repeatedly.
			\item 
			\textbf{Generation of limiting probability distribution:} For a fixed error term $\epsilon$, we apply equation (\ref{stopping _criterion}) to stop quantum walk and generate the limiting probability distribution $\Pi$ mentioned in Definition \ref{limiting_edge_probability_distribution}.
			\item 
			\textbf{Generation of communities} Apply Procedure \ref{Main_procedure} and \ref{refinement_procedure} for generating communities.
		\end{enumerate}
	\end{procedure}

	\section{Numerical experiments}
	\label{Numerical_experiments} 
	
	We consider a number of standard examples of graphs with a community structure, which are considered in literature. Some of them are the random graphs, which may have directed edges or loops. But, our procedure is applicable for undirected simple graphs. Hence, we remove the loops and directed edges if exist. We use NetworkX, a Python library, for all the computations related to graphs \cite{hagberg2007exploring}.
	
	\subsection{Barbell graphs}
	
\begin{figure*}
		\centering
		\begin{subfigure}[b]{0.32\textwidth}
			\centering
			\setlength{\abovecaptionskip}{1pt}
			\setlength{\belowcaptionskip}{0pt}
			
			\includegraphics[width=\linewidth,height=2.6cm]{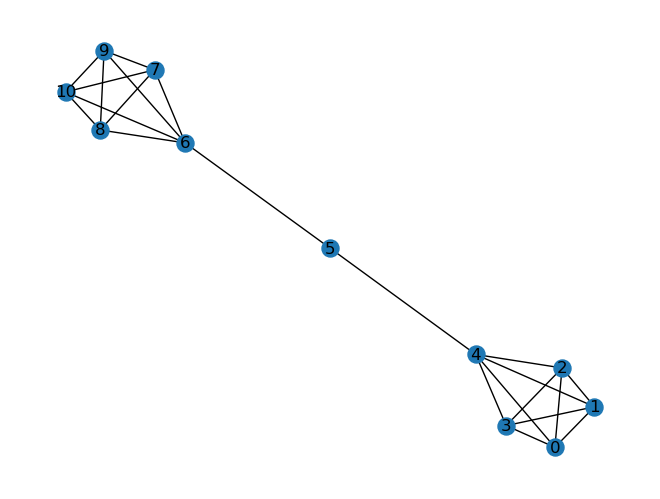}
			\caption{A barbell graph where two complete graph $k_5$ are joined to a central vertex $5$ by two edge $(5, 6), (5, 4)$ .}
			\label{barbell_graph}
		\end{subfigure}
		\hfill
		\begin{subfigure}[b]{0.32\textwidth}
			\centering
			\includegraphics[width=\linewidth,height=2.6cm]{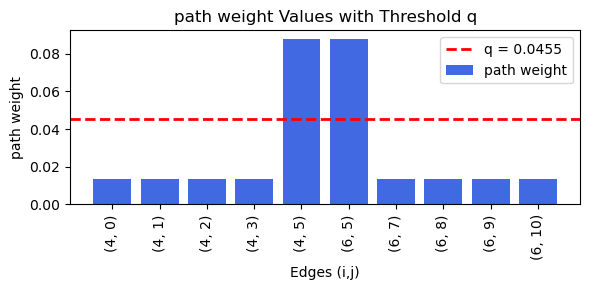}
			\caption{Path weight between different vertices of the for generating the communities from the graph in \ref{barbell_graph}.}
		\end{subfigure}
		\hfill
		\begin{subfigure}[b]{0.32\textwidth}
			\centering
			\includegraphics[width=\linewidth,height=2.6cm]{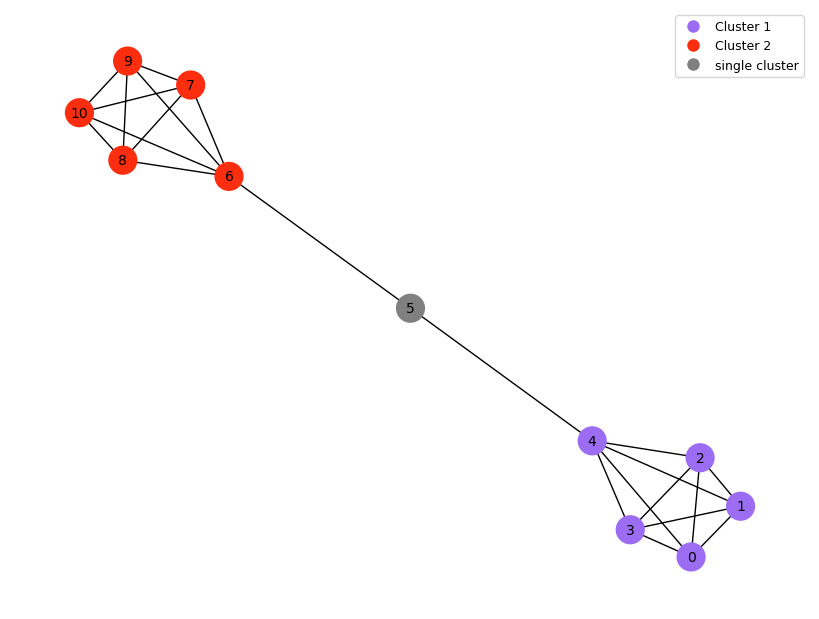}
			\caption{Communities in the graph of Subfigure  \ref{barbell_graph} are extracted with three different colors.}
			\label{Communities_in_berbell_graph}
		\end{subfigure}
		\caption{Community detection on a barbell graph}
		\label{barbell_graph_community}
	\end{figure*}

	In \autoref{barbell_graph}, we consider a barbell graph \cite{ghosh2008minimizing} which is generated by joining two copies of the complete graph $K_5$ with a central vertex by two edges. This graph has eleven vertices and twenty-two edges.
	
	Considering $q = \frac{1}{m} = 0.0455$ in \autoref{combined_procedure}, we obtain three distinct communities in the barbel graph. Two distinct communities arise from two complete graphs. The central vertex forms a community with a single vertex. The coverage values corresponding to $q = 0.0455$  is $0.909$. In \autoref{Communities_in_berbell_graph}, we present the communities generated by \autoref{combined_procedure}.
	
	\subsection{Relaxed caveman graph}
	
\begin{figure*}[t]
		\centering
		\setlength{\abovecaptionskip}{1pt}
		\setlength{\belowcaptionskip}{0pt}
		\begin{subfigure}[b]{0.32\textwidth}
			\centering
			\includegraphics[width=\linewidth,height=2.6cm]{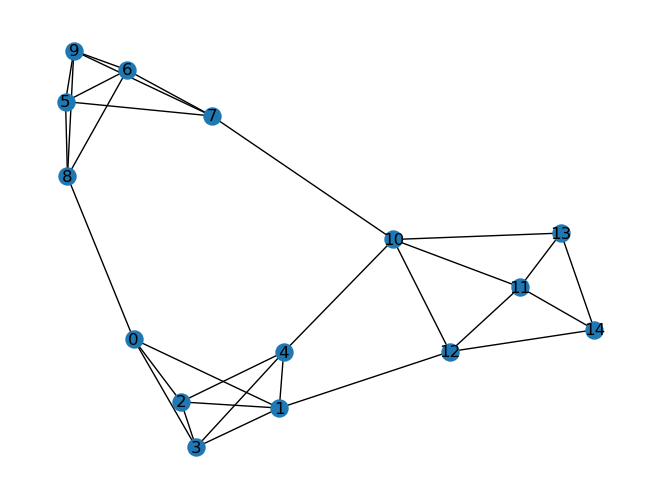}
			\caption{Relaxed caveman graph having three caves with rewiring probability $p = 0.2$.}
			\label{relaxed_caveman_graph}
		\end{subfigure}
		\hfill
		\begin{subfigure}[b]{0.32\textwidth}
			\centering
			\includegraphics[width=\linewidth,height=2.6cm]{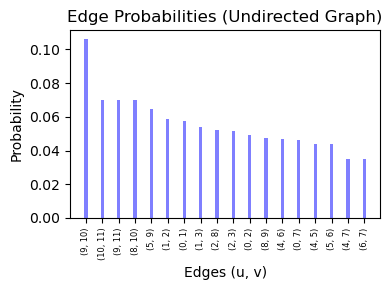}
			\caption{Edge probability in the limiting distribution $\Pi$ is depicted as bar diagram for the graph in Subfigure \ref{relaxed_caveman_graph}.}
		\end{subfigure}
		\hfill
		\begin{subfigure}[b]{0.32\textwidth}
			\centering
			\includegraphics[width=\linewidth,height=2.6cm]{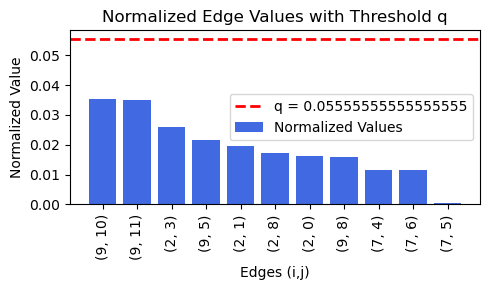}
			\caption{Path weight between different vertices of the for generating the communities from the graph in \ref{relaxed_caveman_graph}.}
		\end{subfigure}
		
		\vspace{1mm}
		
		\begin{subfigure}[b]{0.32\textwidth}
			\centering
			\includegraphics[width=\linewidth,height=2.6cm]{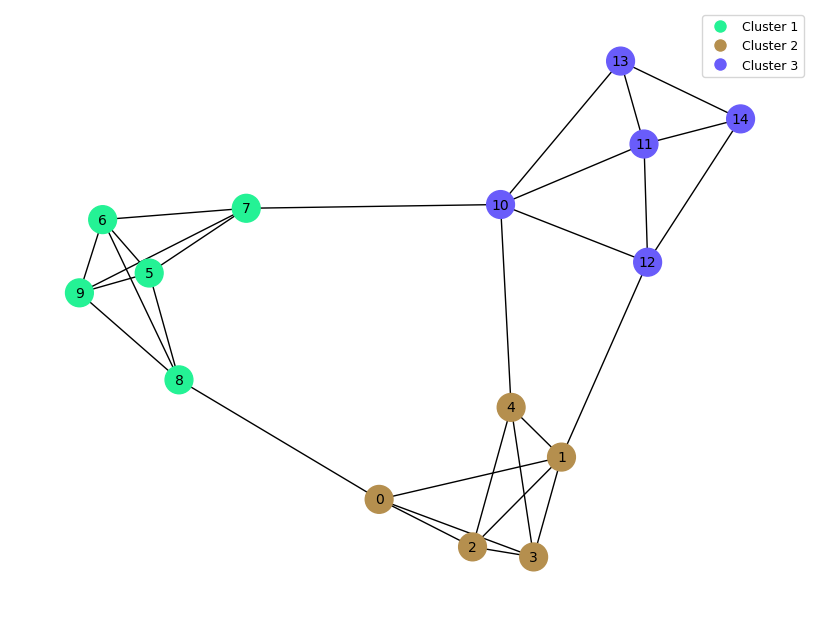}
			\caption{Communities in the graph of Subfigure  \ref{relaxed_caveman_graph} are extracted with three different colors.}
		\end{subfigure}
		\hfill
		\begin{subfigure}[b]{0.32\textwidth}
			\centering
			\includegraphics[width=\linewidth,height=2.6cm]{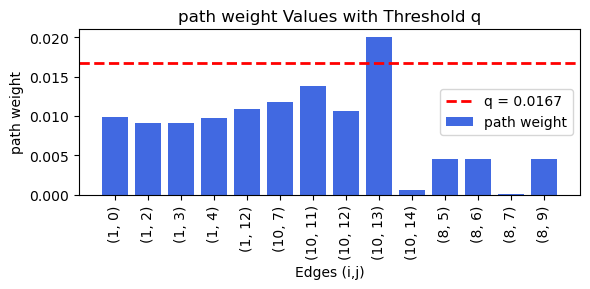}
			\caption{Considering $q = \frac{1}{2m}$, the path weight between different vertices of the graph for generating the communities from the graph in \ref{relaxed_caveman_graph}.}
		\end{subfigure}
		\hfill
		\begin{subfigure}[b]{0.32\textwidth}
			\centering
			\includegraphics[width=\linewidth,height=2.6cm]{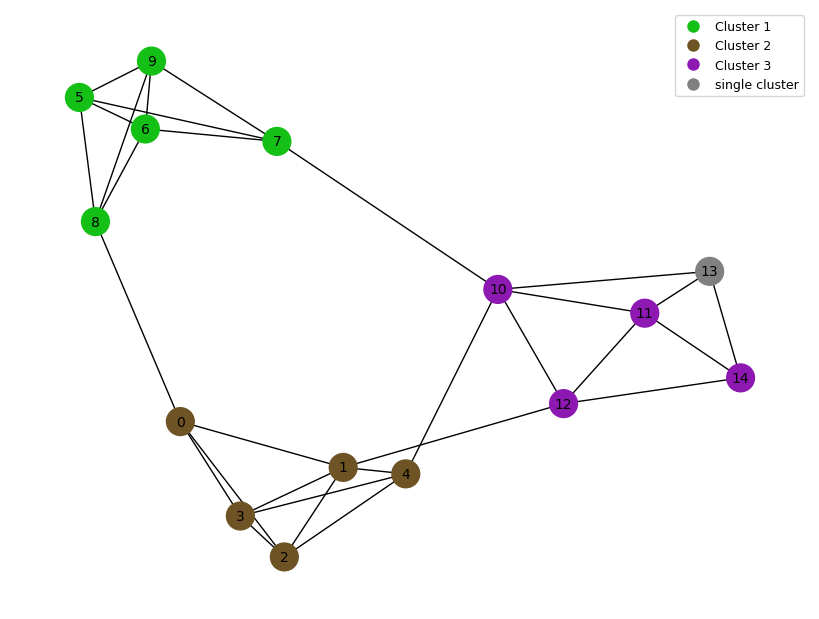}
			\caption{Communities in the graph of Subfigure  \ref{relaxed_caveman_graph} are extracted with four different colors.} 
		\end{subfigure}
		
		\caption{Community detection in a relaxed caveman graph.}
		\label{community_in_relaxed_caveman_graph}
		
	\end{figure*}

	The connected caveman graph \cite{watts1999small, watts1999networks} is generated by joining complete subgraphs each with $k$ vertices. An edge in a clique is rewired to a node in an adjacent clique with probability $p$. In \autoref{relaxed_caveman_graph} we depict a relaxed caveman graph having $30$ edges, and $15$ vertices which are equally distributed in $3$ caves. We assume $p = 0.2$ which creates $4$ inter-community edges.
	
	Putting $q = \frac{1}{m} = 0.033$, we find three communities which are presented in Sub-figure \ref{community_in_relaxed_caveman_graph}. Decreasing the value of $q_2$ to $0.0166$ we find more communities in our network. Coverage are $0.866$ and $0.766$ for $q = 0.0333$ and $0.0166$, respectively.
	
	\subsection{Planted $l$-partition graph}
	
\begin{figure*}
		\centering
		\begin{subfigure}[b]{0.32\textwidth}
			\centering
			\setlength{\abovecaptionskip}{1pt}
			\setlength{\belowcaptionskip}{0pt}
			
			\includegraphics[width=\linewidth,height=2.6cm]{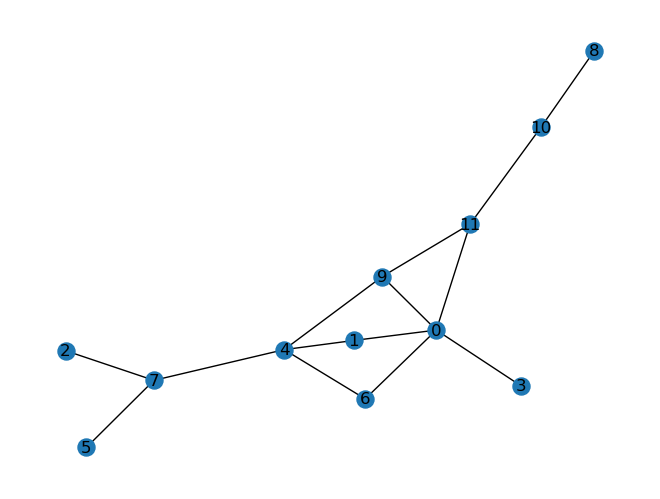}
			\caption{A planted partition graph with $p_1 = 0.4$ and $p_2 = 0.1$.}
			\label{planted_partition_graph}
		\end{subfigure}
		\hfill
		\begin{subfigure}[b]{0.32\textwidth}
			\centering
			\includegraphics[width=\linewidth,height=2.6cm]{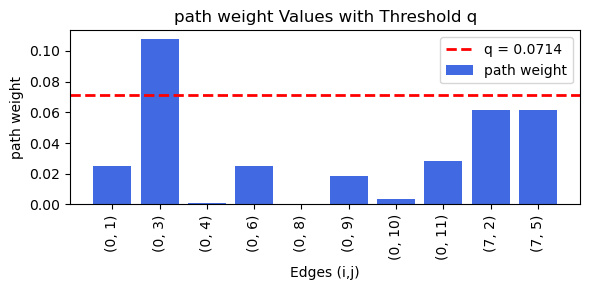}
			\caption{Path weight between different vertices of the for generating the communities from the graph in \ref{planted_partition_graph}.}
		\end{subfigure}
		\hfill
		\begin{subfigure}[b]{0.32\textwidth}
			\centering
			\includegraphics[width=\linewidth,height=2.6cm]{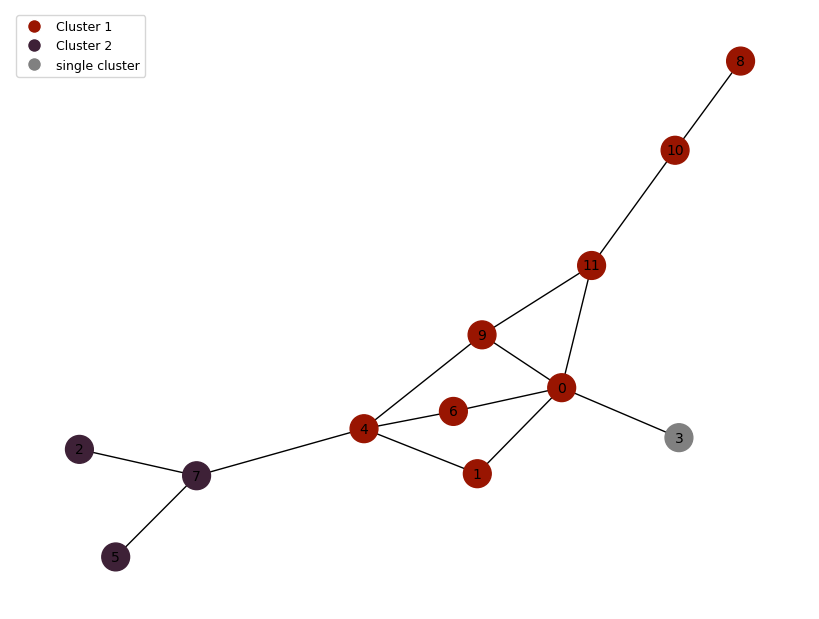}
			\caption{Communities in the graph of Subfigure  \ref{planted_partition_graph} are extracted with three different colors.}
		\end{subfigure}
		\caption{Community detection in a planted partition graph.}
		\label{planted_partition_graph_community}
	\end{figure*}

	The planted $l$-partition graph \cite{fortunato2010community, condon2001algorithms} has $n = l \times k$ vertices distributed into $l$ communities with $k$ vertices in each. Let $p$ and $q$ with $p > q$ be the probability of existence of the intra-community and inter-community edges, respectively. \autoref{planted_partition_graph} contains a planted $l$-partition graph with $14$ edges and $12$ vertices which are distributed into $3$ communities each having $4$ vertices in it. Also, $p_1 = 0.4$ and $p_2 = 0.1$, respectively.
	
	One of the highest degree vertex $0$ is randomly selected as the first vertex to initiate \autoref{Main_procedure}. Considering $q > 0.071$, we obtain three communities, where a community contain a single vertex. The coverage values corresponding to $q = 0.071$ is $0.857$.
	
	\subsection{Zachary's karate club graph}
	
	The Zachary's karate-club network \cite{zachary1977information} is another well-studied graph considered in the literature of community detection \cite{zhang2013finding, mukai2020discrete}. It has $34$ vertices and $78$ edges. Article \cite{zhang2013finding} presents three communities in it. In contrast, article \cite{mukai2020discrete} consider the same network that found two communities only. 
	
\begin{figure*}
		\centering
		\begin{subfigure}[b]{0.32\textwidth}
			\centering
			\setlength{\abovecaptionskip}{1pt}
			\setlength{\belowcaptionskip}{0pt}
			
			\includegraphics[width=\linewidth,height=2.6cm]{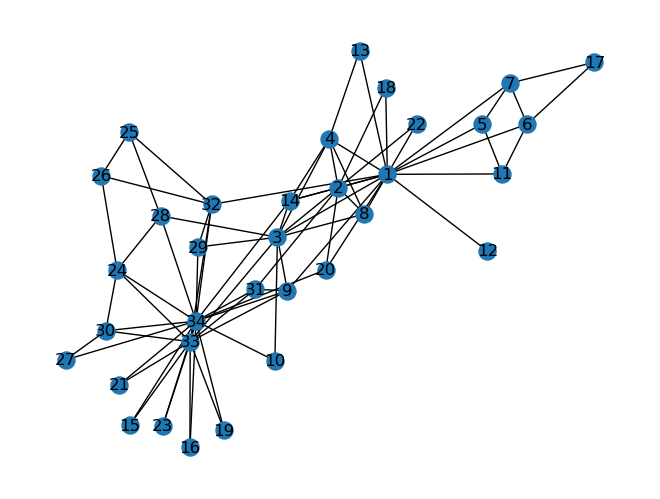}
			\caption{Zachary's karate club network.}
			\label{karate_club_graph}
		\end{subfigure}
		\hfill
		\begin{subfigure}[b]{0.32\textwidth}
			\centering
			\includegraphics[width=\linewidth,height=2.6cm]{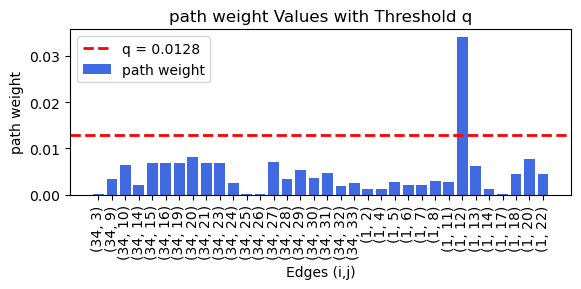}
			\caption{In \ref{karate_club_graph}, the path weight between various vertices of the graph is used to generate the communities.}
		\end{subfigure}
		\hfill
		\begin{subfigure}[b]{0.32\textwidth}
			\centering
			\includegraphics[width=\linewidth,height=2.6cm]{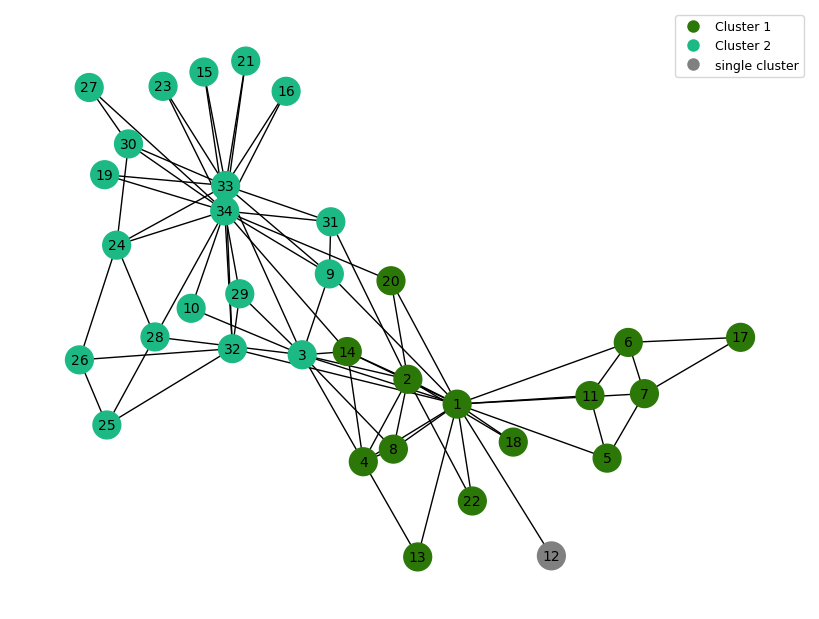}
			\caption{Three distinct colors are used to extract communities from the graph in Subfigure \ref{karate_club_graph}.}
			\label{communities_in_karate_club_graph}
		\end{subfigure}
		\caption{Karate club graph are considered to generate communities.}
		\label{karate club graph}
	\end{figure*}

	Considering $q = \frac{1}{m} = 0.0128$ in \autoref{combined_procedure}, we obtain three communities in this network. Two communities have fourteen and eighteen vertices, and one community consists of a single vertex. The coverage values corresponding to $q = 0.0128$  is $0.859$. In \autoref{communities_in_karate_club_graph}, we present the communities determined by \autoref{combined_procedure}.
	
	\subsection{Les Mis\'{e}rable graph}
	
\begin{figure*}
		\centering
		\begin{subfigure}[b]{0.32\textwidth}
			\centering
			\setlength{\abovecaptionskip}{1pt}
			\setlength{\belowcaptionskip}{0pt}
			
			\includegraphics[width=\linewidth,height=2.6cm]{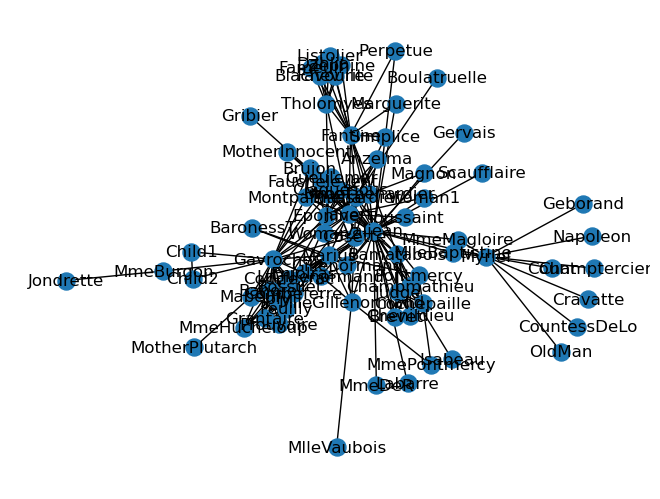}
			\caption{Les Mis\'{e}rable graph}
			\label{Les_Miserable_graph}
		\end{subfigure}
		\hfill
		\begin{subfigure}[b]{0.32\textwidth}
			\centering
			\includegraphics[width=\linewidth,height=2.6cm]{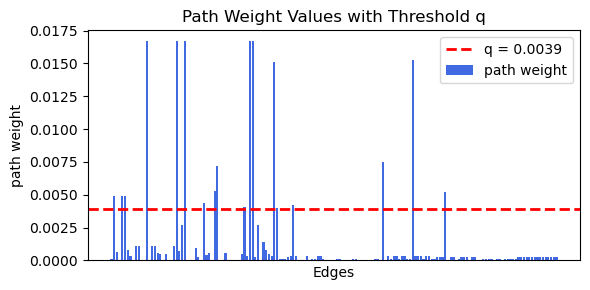}
			\caption{In \ref{Les_Miserable_graph}, the path weight between various vertices of the graph is used to generate the communities.}
		\end{subfigure}
		\hfill
		\begin{subfigure}[b]{0.32\textwidth}
			\centering
			\includegraphics[width=\linewidth,height=2.6cm]{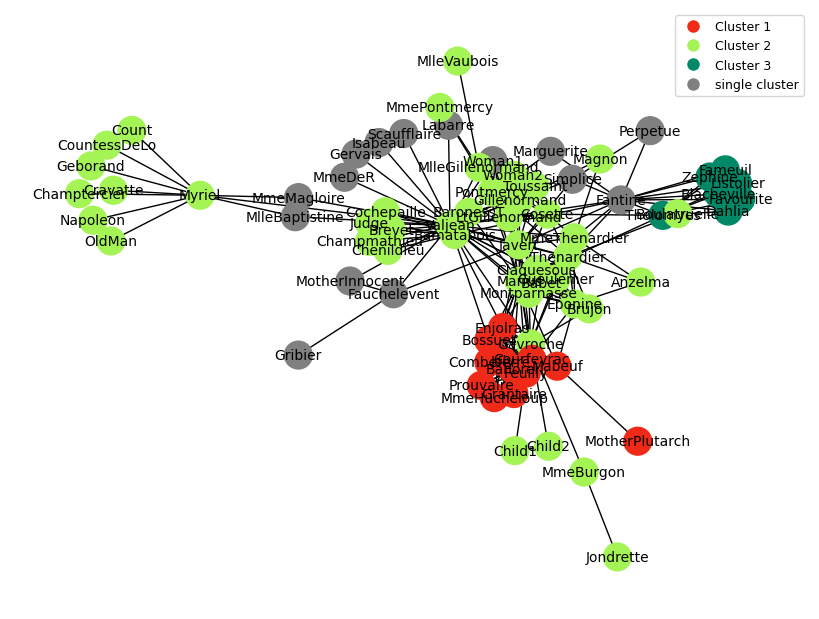}
			\caption{The three communities are represented by three distinct colors in the graph shown in Subfigure \ref{Les_Miserable_graph}, while all other single-node clusters are shown by gray.}
			\label{communities_in_Les_Miserable_graph}
		\end{subfigure}
		\caption{Les Misérable graph are utilized to generate communities.}
		\label{Les Misérable graph}
	\end{figure*}

	\autoref{Les_Miserable_graph} represents the ``Les Mis\'{e}rable" graph \cite{knuth1993stanford}. having $77$ vertices and $254$ edges.
	
	Article \cite{zhang2013finding} determines six communities in the graph. We use \autoref{combined_procedure} considering $q = \frac{1}{m} = 0.0039$ for determining the communities, that generates eighteen communities. Among them there are three communities with $8$, $12$, and $44$ vertices. Remaining $15$ communities have single vertices in each. In \autoref{communities_in_Les_Miserable_graph}, we present these communities. The coverage values corresponding to $q = 0.0039$ is $0.752$.
	
	\subsection{Dolphin's social network}
	
\begin{figure*}
		\centering
		\begin{subfigure}[b]{0.32\textwidth}
			\centering
			\setlength{\abovecaptionskip}{1pt}
			\setlength{\belowcaptionskip}{0pt}
			
			\includegraphics[width=\linewidth,height=2.6cm]{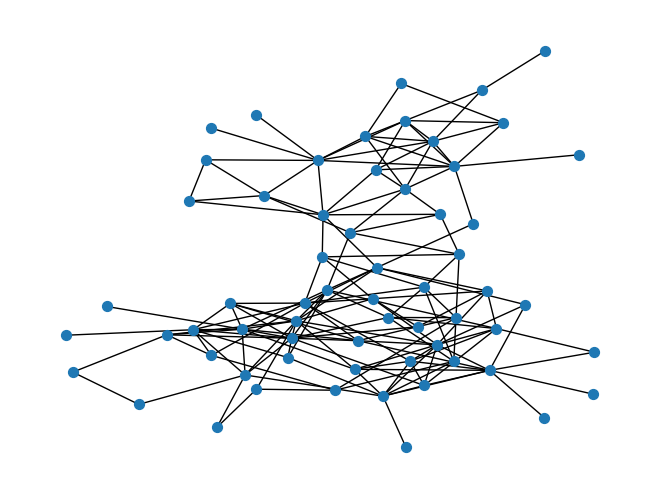}
			\caption{Dolphin social network}
			\label{dolphins_Graph}
		\end{subfigure}
		\hfill
		\begin{subfigure}[b]{0.32\textwidth}
			\centering
			\includegraphics[width=\linewidth,height=2.6cm]{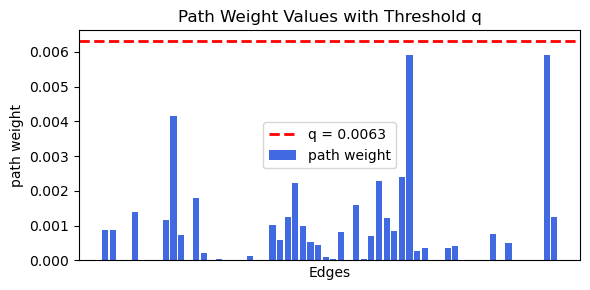}
			\caption{In \ref{dolphins_Graph}, the path weight between various vertices of the graph is used to generate the communities.}
		\end{subfigure}
		\hfill
		\begin{subfigure}[b]{0.32\textwidth}
			\centering
			\includegraphics[width=\linewidth,height=2.6cm]{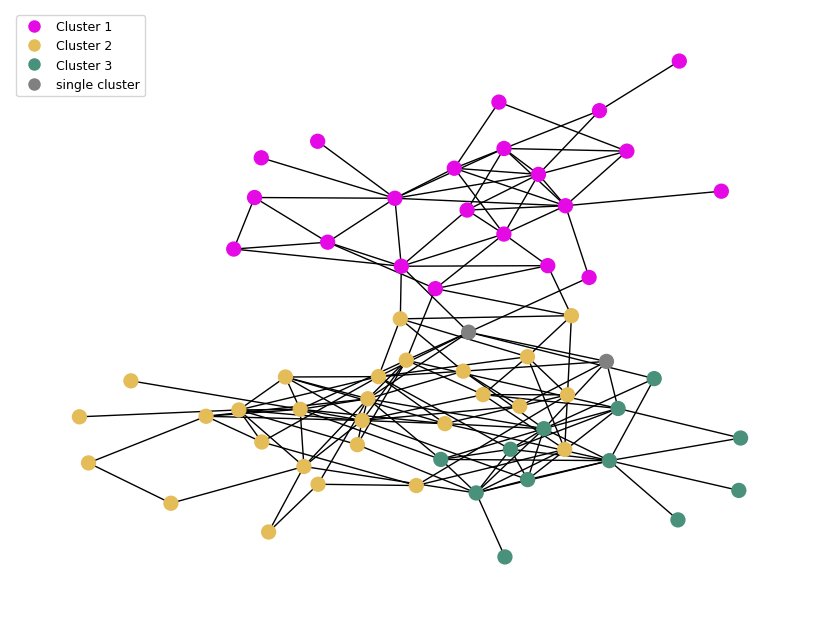}
			\caption{The three communities are represented by three distinct colors in the graph shown in Subfigure \ref{dolphins_Graph}, while the other two single-node communities are shown by gray.}
		\end{subfigure}
		\caption{Dolphin Network are utilized to generate communities.}
		\label{dolphins Graph}
	\end{figure*}

	\autoref{dolphins_Graph} illustrates the dolphin's social network \cite{lusseau2003bottlenose}. Article \cite{lu2012community} determines four communities in the network, that consists of $6$, $13$, $22$, and $21$ vertices. 
	
	Considering $q = \frac{1}{m} = 0.0063$ in \autoref{combined_procedure}, we obtain $5$ communities. There are three communities having $12$, $21$, and $27$ vertices, and two communities consists of a single vertex in each. These single-vertex communities exist because the respective vertices are associated with several communities and hence cannot be unequivocally allocated to any one of the three communities. The coverage values corresponding to $q = 0.0063$  is $0.824$. \autoref{dolphins Graph} presents the communities generated by our procedure. 
	
	\section{Conclusion}
	
	In this article, we introduce a new procedure that utilizes Szegedy quantum walks to detect the communities in a network. There are multiple models of quantum walks in the literature. We employ the Szegedy quantum walks on graphs for our investigation.
	
	The limiting probability distribution of the Szegedy quantum walks assists us in finding the communities. This distribution depends on the choice of the initial state of the quantum walker. We assume that the walker begins its journey from the vertices with the highest degree. The initial quantum state as well as the limiting probability distribution depends on the structure of the graph. We use it in \autoref{combined_procedure} to extract the communities from the graph. This procedure is a combination of two subroutines, which are \autoref{Main_procedure} and \autoref{refinement_procedure}. In \autoref{Main_procedure}, we find the communities using quantum walks. Then, \autoref{refinement_procedure} refines the preliminary communities. We apply \autoref{combined_procedure} on different graphs to identify the communities in them. 
	
	In the theory of complex network, the problem of community detection is an important problem, because the communities assists us to understand the structure of the networks. In the future, the interested reader may apply our community detection procedure for investigating the structure of  different complex networks.
	
	\section*{Acknowledgment}
	
	MSR is thankful to Ministry of Education, Government of India for a doctoral fellowship. 
	
	\section*{Data availability statement} 
	
	No dataset if generated during this work. All the programs to plot the figures are available in GitHub \url{https://github.com/dosupriyo/Community-detection-using-Szegedy-quantum-walk.git}.
	
	\appendix
	\section{Dependency on the initial state}
	\label{Dependency_on_initial_state} 
	
	\begin{figure}[t]
		\centering
		\setlength{\abovecaptionskip}{2pt}
		\setlength{\belowcaptionskip}{0pt}
		
		\begin{subfigure}[b]{0.48\linewidth}
			\centering
			\scalebox{0.8}{
				\begin{tikzpicture}[every node/.style={circle, draw, minimum size=4.5mm, font=\scriptsize}]
					\node[fill=black!20] (0) at (-1,0) {0};
					\node[fill=black!20] (1) at (0,0) {1};
					\node[fill=black!20] (2) at (-2,0) {2};
					\node[fill=black!20] (3) at (1,1.5) {3};
					\node[fill=black!20] (4) at (1,-1.5) {4};
					\node[fill=black!20] (5) at (-3,1.5) {5};
					\node[fill=black!20] (6) at (-3,-1.5) {6};
					
					\draw (0)--(1);
					\draw (0)--(2);
					\draw (1)--(3);
					\draw (1)--(4);
					\draw (3)--(4);
					\draw (2)--(5);
					\draw (2)--(6);
					\draw (5)--(6);
			\end{tikzpicture}}
			\caption{A barbell graph.}
			\label{initial_state_graph_example}
		\end{subfigure}
		\hfill
		\begin{subfigure}[b]{0.48\linewidth}
			\centering
			\scriptsize
			\setlength{\tabcolsep}{2.5pt}
			\renewcommand{\arraystretch}{1.1}
			\begin{tabular}{|c|c|c|}
				\hline
				Edge & $\Pi'$ & $\Pi$ \\
				\hline
				(0,1) & 0.1274 & 0.1625 \\
				(0,2) & 0.1274 & 0.1625 \\
				(1,3) & 0.1274 & 0.1131 \\
				(1,4) & 0.1274 & 0.1131 \\
				(2,5) & 0.1274 & 0.1131 \\
				(2,6) & 0.1274 & 0.1131 \\
				(3,4) & 0.1274 & 0.1116 \\
				(5,6) & 0.1274 & 0.1116 \\
				\hline
			\end{tabular}
			\caption{Limiting distributions.}
			\label{different_probability_distributions}
		\end{subfigure}
		\caption{A barbell graph and different limiting probability distributions for different initial states.}
	\end{figure}

	We consider an initial state, mentioned in equation \eqref{initial_state}. Consider another state
	\begin{equation}
		\label{equal_superposition_state}
		\ket{\psi_0'} = \frac{1}{\sqrt{|E(G)|}} \sum_{i \in V(G)}\ket{\overrightarrow{(i,j)}}.
	\end{equation}
	Different choice of initial states makes the limiting probability distribution different. For example consider the barbell graph depicted in \autoref{initial_state_graph_example}. The state in equation \eqref{equal_superposition_state} is an eigenvector of the evolution operator $U_{sz}$. Therefore, the probability distribution on edges $\Pi'$ generated by $\ket{\psi_0'}$ is unchanged in every time step. But, choosing the initial state $\ket{\psi_0}$ mentioned in equation \eqref{initial_state} we obtain probability distribution $\Pi$. We observe two edges $(0, 1)$ and $(0, 2)$ having higher probability values than the others. Removing them, we get two communities. We mention the two probability distributions $\Pi$ and $\Pi'$ in Table \ref{different_probability_distributions}.
	
\begin{figure}
		
		\begin{subfigure}{0.45\textwidth}
			\centering
			\includegraphics[width=7cm,height=4cm]{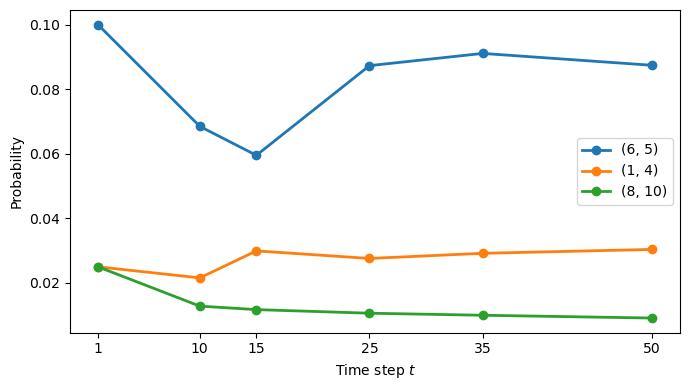}
			\caption{Convergence of edge probability values over time to the limiting distribution $\Pi$ for a sample of edges of the barbell graph, depicted in \autoref{barbell_graph}. We observe the convergence from $T = 743$.}
		\end{subfigure}
		\hfill
		\begin{subfigure}{0.45\textwidth}
			\centering
			\includegraphics[width=7cm,height=4cm]{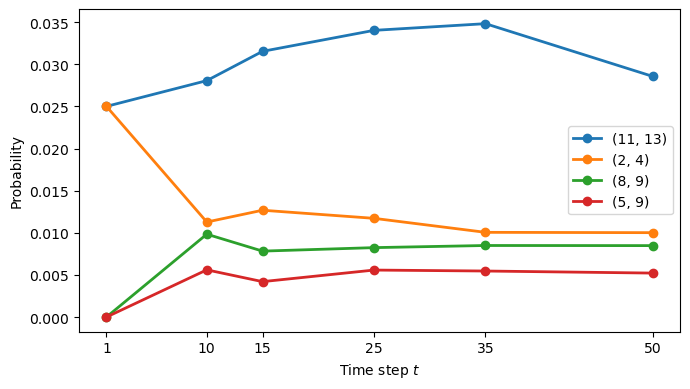}
			\caption{Convergence of edge probability values over time to the limiting distribution $\Pi$ for a sample of edges of the relaxed caveman graph, depicted in \autoref{relaxed_caveman_graph}. We observe the convergence from $T = 1329$.} 
		\end{subfigure}
		
		\vspace{0.3cm} 
		
		\begin{subfigure}{0.45\textwidth}
			\centering
			\includegraphics[width=7cm,height=4cm]{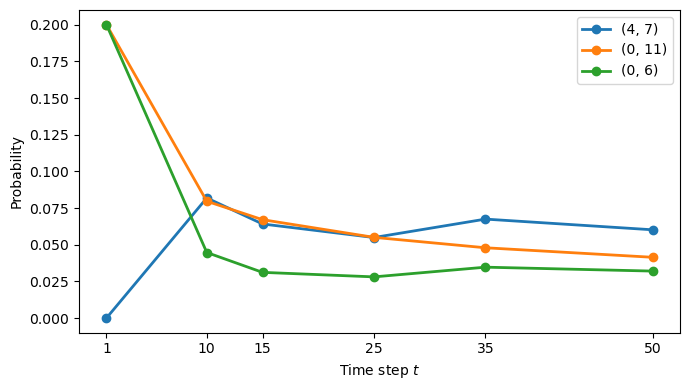}
			\caption{Convergence of edge probability values over time to the limiting distribution $\Pi$ for a sample of edges of the planted partition graph, depicted in \autoref{planted_partition_graph}. We observe the convergence from $T = 1094$.}
		\end{subfigure}
		\hfill
		\begin{subfigure}{0.45\textwidth}
			\centering
			\includegraphics[width=7cm,height=4cm]{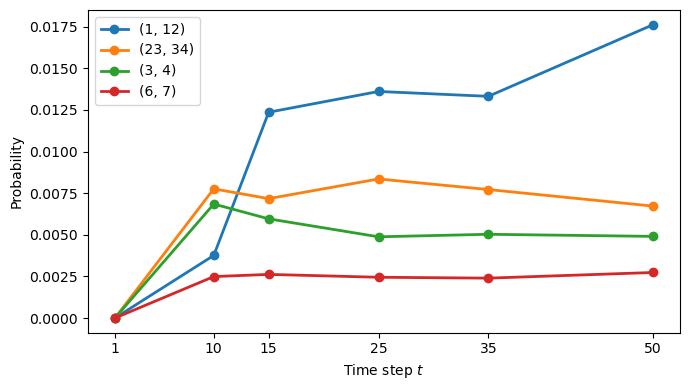}
			\caption{Convergence of edge probability values over time to the limiting distribution $\Pi$ for a sample of edges of the karate club graph, depicted in \autoref{karate_club_graph}. We observe the convergence from $T = 882$.}
		\end{subfigure}
		
		\vspace{0.3cm} 
		
		\begin{subfigure}{0.45\textwidth}
			\centering
			\includegraphics[width=7cm,height=4cm]{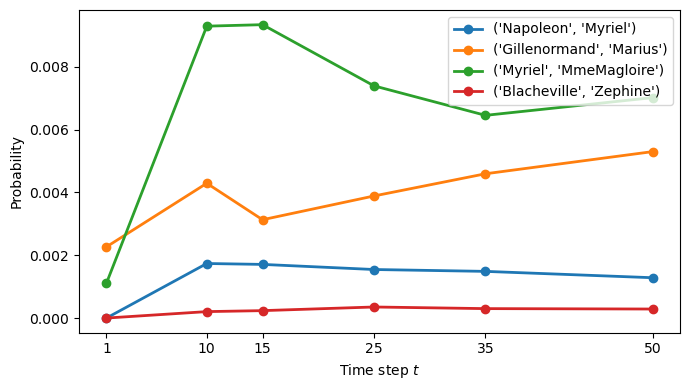}
			\caption{Convergence of edge probability values over time to the limiting distribution $\Pi$ for a sample of edges of the Les Mis\'{e}rable graph, depicted in \autoref{Les_Miserable_graph}. We observe the convergence from $T = 714$.} 
		\end{subfigure}
		\hfill
		\begin{subfigure}{0.45\textwidth}
			\centering
			\includegraphics[width=7cm,height=4cm]{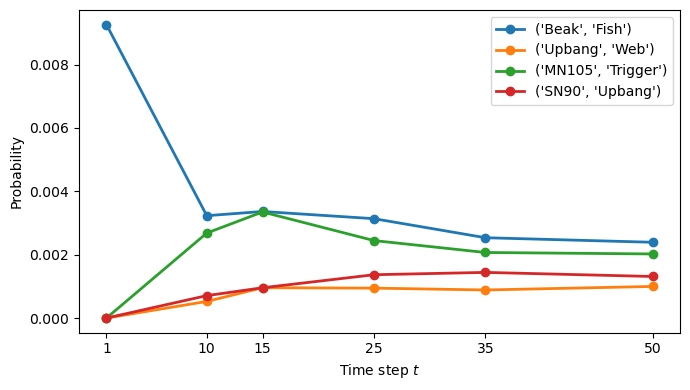}
			\caption{Convergence of edge probability values over time to the limiting distribution $\Pi$ for a sample of edges of the dolphins Graph, depicted in \autoref{dolphins_Graph}. We observe the convergence from $T = 705$.}
		\end{subfigure}

		\caption{Convergence of edge probability distributions, which is mentioned in \autoref{limiting_edge_probability_distribution}.}
		\label{convergence_of_probability}
	\end{figure}

	
	The edge probability vectors, mentioned in \autoref{edge_probability_vectors}, converges to a limiting distribution, mentioned in \autoref{limiting_edge_probability_distribution}, which we can observe numerically. We consider a random sample of $4$ edges $\{(1, 12), (23, 24), (3, 4), (6, 7)\}$ from the karate club graph, which is depicted in \autoref{karate_club_graph}. As none of them is incident to the highest degree vertices their probability values at $t = 0$ is $0$. Upto $t = 50$ we present the changes in probability values which is depicted in \autoref{convergence_of_probability}.
	
\end{document}